\documentclass[graybox]{svmult}

\usepackage{cite}
\usepackage{mathptmx}
\usepackage{helvet}
\usepackage{courier}

\usepackage{makeidx}
\usepackage{graphicx}
\usepackage{multicol}
\usepackage[bottom]{footmisc}

\usepackage{amsmath}
\usepackage{braket}
\usepackage{caption}
\usepackage{subfig}

\newcommand{\PT}{\ensuremath{\mathcal{PT}}}

\makeindex

\begin{document}
	
\title*{Sublattice signatures of transitions in a \PT-symmetric dimer lattice}
\author{Andrew K. Harter and Yogesh N. Joglekar}
\institute{Andrew K. Harter \at Indiana University Purdue University Indianapolis(IUPUI), Indianapolis, Indiana 46202 USA, \email{akharter@iupui.edu}
	\and Yogesh N. Joglekar \at Indiana University Purdue University Indianapolis (IUPUI), Indianapolis, Indiana 46202 USA, \email{yojoglek@iupui.edu}}
\maketitle

\abstract*{}

\abstract{Lattice models with non-hermitian, parity and time-reversal (\PT) symmetric Hamiltonians, realized most readily in coupled optical systems, have been intensely studied in the past few years. A \PT-symmetric dimer lattice consists of dimers with intra-dimer coupling $\nu$, inter-dimer coupling $\nu'$, and balanced gain and loss potentials $\pm i\gamma$ within each dimer. This model undergoes two independent transitions, namely a \PT-breaking transition and a topological transition. We numerically and analytically investigate the signatures of these transitions in the time-evolution of states that are initially localized on the gain-site or the loss-site.}

%--------------------------------------------------------------------------------------%

\section{Introduction}
\label{sec:1}
Finite, discrete systems have always been an important testing ground in that they are often amenable to straightforward numerical approach, while retaining the complex and interesting features of their infinite and continuum counterparts. Lattice models, where a quantum particle occupies discrete locations and only tunnels between adjacent sites, successfully describe physical properties of a number of crystalline, condensed matter systems~\cite{ashmermin,cardona} as well as light propagation in arrays of coupled optical waveguides~\cite{waveguides} in the paraxial approximation~\cite{yariv}. A dimer model, where the tunneling strength alternates between two values, was first explored by Su, Schrieffer, and Heeger (SSH) in the context of solitons in polyacetylene~\cite{ssh1,ssh2}. Since then, the one-dimensional SSH model has been extensively studied because it exhibits topologically non-trivial edge states~\cite{aahdassarma} and its generalizations lead to band structures with nonzero Chern numbers~\cite{suchen1,suchen2}. 

Realizations of an SSH model in coupled optical waveguides instead of the nature-given long acetylene chains are advantageous~\cite{review}. In the former, the ratio of tunneling strengths, and the size and parity of the dimer chain can be varied over a wide range, and the entire bandwidth of the SSH band structure is accessible;and one can model non-hermitian, gain and loss potentials because  the absorption and amplification of electromagnetic waves are both easily implemented~\cite{henning}. Experimental realizations of non-uniform waveguide lattices have been demonstrated with lattice sites $N\sim10-100$~\cite{cs}, single-site or wide-beam input~\cite{ys}, and single-photon source inputs~\cite{job}; in particular, edge states and their adiabatic transfer in quasi-periodic waveguide lattices have been experimentally investigated~\cite{top}. 

The past five years have seen a surge of interest, driven primarily by experiments on optical systems~\cite{expt1,expt2,expt3,uni1,uni2,expt4,expt5,expt6,expt7,expt8,expt9}, in open systems that are faithfully described by an effective,  non-hermitian Hamiltonian that is invariant under combined parity and time-reversal (\PT) operations~\cite{bender1,bender2}. Typically, a \PT-symmetric Hamiltonian $H$ is comprised of a hermitian, kinetic energy term $H_0$ and a non-hermitian, \PT-symmetric potential term $V=\PT V\PT\neq V^\dagger$ that represents balanced, spatially separated gain and loss. Although $H$ is not hermitian, its spectrum is purely real when the strength of the non-hermitian potential is small, and changes into complex-conjugate pairs when it exceeds a threshold called the \PT-breaking threshold~\cite{bender2}. In contrast with the traditional hermitian case, the non-hermitian, \PT-symmetric Hamiltonian is defective at the \PT-breaking threshold~\cite{kato,ingrid}. A \PT-symmetric SSH model, or equivalently a dimer model, has gain and loss of equal strengths on alternate sites~\cite{kottos}, and is mathematically equivalent to a dimer model which has only a loss term on every other site. This purely lossy dimer model shows a quantized mean displacement that, under certain constraints, has a topological origin. This transition is driven by the ratio of inter-dimer and intra-dimer tunneling amplitudes, and befitting a topological transition, is independent of the strength of the loss potential and robust over a broad range of model parameters~\cite{lr}. 

In this paper, we discuss the properties of \PT-symmetric dimer model over a wide range of parameters, such that it undergoes {\it both the \PT-breaking transition and the topological transition.} The \PT-breaking transition in a \PT-dimer model was studied by Zheng {\it et al.}~\cite{kottos}, and the topological transition in a purely lossy dimer model was predicted by Rudner and Levitor~\cite{lr}. Neither, however, investigated the interplay between these two transitions. 

The plan of the paper is as follows. In Sect.~\ref{sec:2} we present the key properties of a \PT-symmetric dimer model, as they relate to the two transitions it undergoes. In Sect.~\ref{sec:3} we present numerical results for the time evolution of a wave packet that is initially localized on the gain site or a loss site. Since the intensities on the gain sites are orders of magnitude higher than those on the loss sites, particularly in the \PT-broken phase, we separately consider the intensity distributions on the gain-sublattice and the loss-sublattice. We show that these distributions undergo a qualitative change across the topological transition. In Sect.~\ref{sec:4}, we obtain approximate, analytical expressions for the two sublattice intensity distributions. We conclude the paper with a brief discussion in Sect.~\ref{sec:5}. Our results show that the signatures of the topological transition imprint themselves on the sublattice intensity distributions in the broken \PT-symmetric phase.

%--------------------------------------------------------------------------------------%

\section{The \PT-symmetric dimer model}
\label{sec:2}
\begin{figure}
\centering
\includegraphics[width=0.75\textwidth]{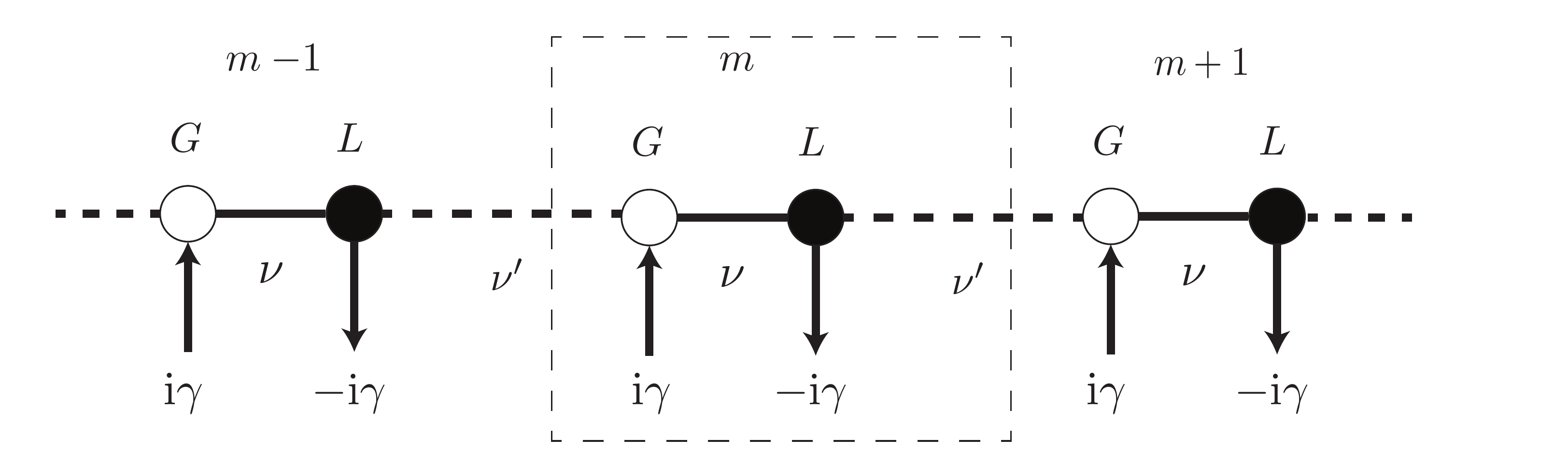}
\caption{Schematic of a \PT-symmetric dimer lattice. The gain-sites $G$, shown by open circles, have a gain potential $+i\gamma$ while the loss sites $L$, shown by black solid circles, have the decay potential $-i\gamma$. The dashed rectangular box indicates the central, $m=0$, dimer. The tunneling within a dimer is given by $\nu$ and the inter-dimer tunneling is $\nu'$.}
\label{fig:ptdimerlattice}
\end{figure}

In this section, we establish the notation and recall results for the \PT-breaking transition in a dimer lattice~\cite{kottos}, and the topological transition in a purely lossy dimer lattice~\cite{lr}. Let us consider a \PT-symmetric dimer lattice, where each dimer consists of a gain site ($G$) with potential $+i\gamma$ and a loss site $L$ with potential $-i\gamma$. The dimer is labeled by the index $m$ where $-M\leq m\leq M$ denotes a finite lattice with $N=2M+1$ dimers, $\nu$ denotes the tunneling within a dimer, and $\nu'$ denotes the tunneling between two adjacent dimers. In this case, the parity operator $\mathcal{P}$ exchanges the gain and the loss sites {\it within each dimer} whereas the time-reversal operator $\mathcal{T}$ corresponds to complex conjugation, thus interchanging the gain with the loss. Figure~\ref{fig:ptdimerlattice} shows a schematic of such a lattice. 

The non-hermitian, \PT-symmetric Hamiltonian $H=H_0+V$ for the lattice is given by 
\begin{eqnarray}
\label{eq:h0}
H_0& =& -\sum_{m=-M}^{M-1} \left(\nu |mG\rangle\langle mL| + \nu' |mL\rangle\langle m+1 G|+\mathrm{h.c.}\right),\\
\label{eq:v}
V& = & +i\gamma\sum_{m=-M}^M\left( |mG\rangle\langle mG|-|mL\rangle\langle mL|\right),
\end{eqnarray}
where $|mG\rangle$ and $|mL\rangle$ denote single-particle states localized on the gain and loss sites of dimer $m$, h.c. denotes the hermitian conjugate, and we have considered a lattice with open boundary conditions. In the Fourier space, this Hamiltonian is block-diagonalized into 2$\times$2 sectors given by
\begin{equation}
\label{eq:hk}
H_{k_n}=\left[\begin{array}{cc} i\gamma & -\nu_{k_n}^* \\ -\nu_{k_n} & -i\gamma\end{array}\right]=i\gamma\sigma_z-(\nu+\nu'\cos k_n)\sigma_x-\nu'\sin k_n\sigma_y. 
\end{equation}
Here $\sigma_i$ are the Pauli matrices, $\nu_{k_n}=\nu+\nu'\exp(ik_n)$, * denote complex conjugation, and $k_n=n\pi/(N+1)$ ($1\leq n\leq N$) are the eigenmomenta consistent with open boundary conditions. For periodic boundary conditions, the corresponding eigenmomenta are given by $k_n=2\pi n/N$ with $|n|\leq (N/2)$. The spectrum of the Hamiltonian $H_k$ is given by $\pm\epsilon_k=\pm\sqrt(|\nu_k|^2-\gamma^2)$; therefore, the \PT-breaking threshold for the dimer lattice is given by $\gamma_{PT}=\min_k(|\nu_k|)$ and becomes, in the infinite-lattice limit~\cite{kottos}, $\gamma_{PT}=|\nu_k|_{k=\pi}=|\nu-\nu'|$.

% G-L sublattice distributions.
\begin{figure}
\centering
\subfloat[][Gain sublattice intensity profile]{\includegraphics[width=0.48\textwidth]{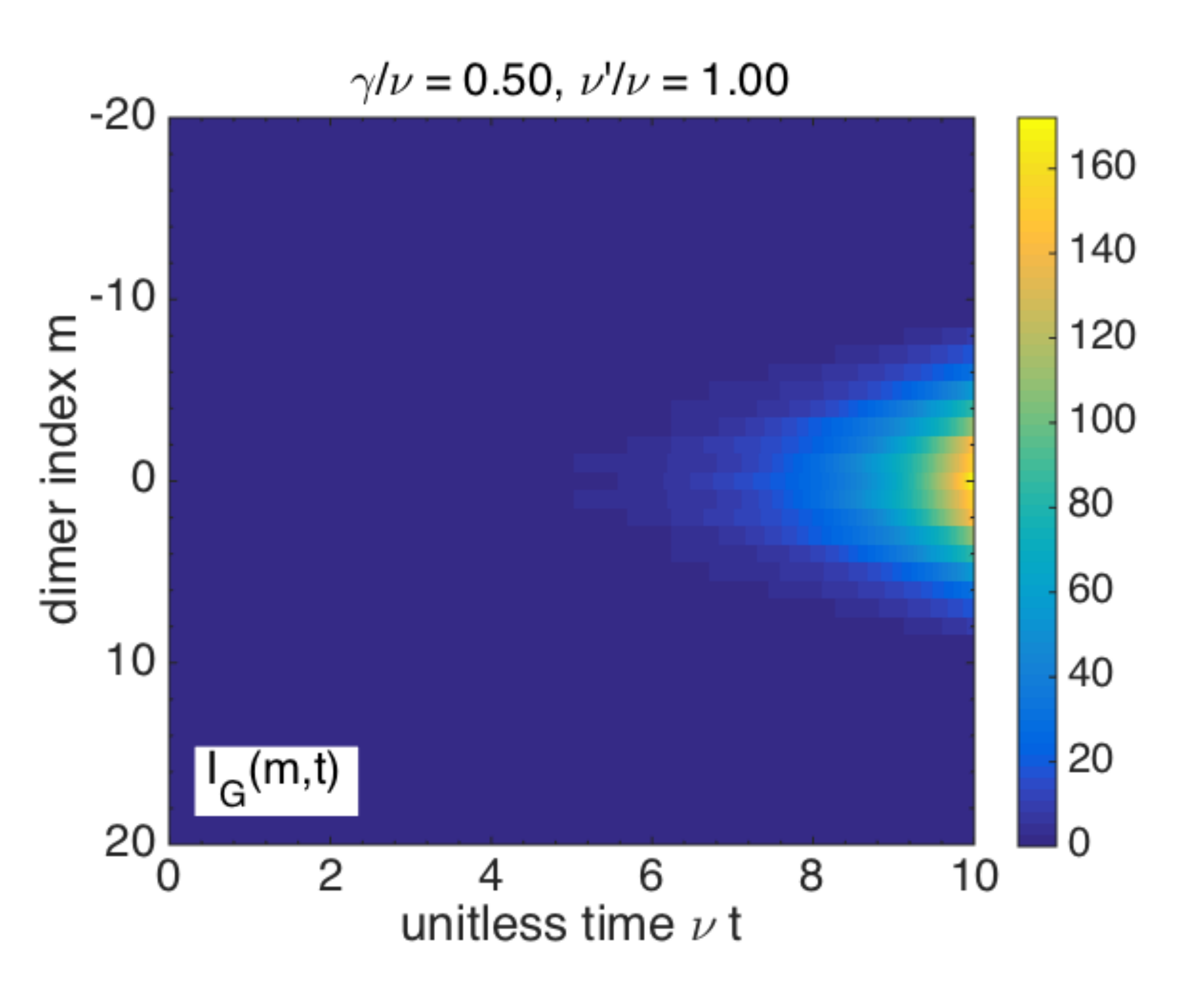}
\label{fig:G-51}}
\subfloat[][Loss sublattice intensity profile]{\includegraphics[width=0.48\textwidth]{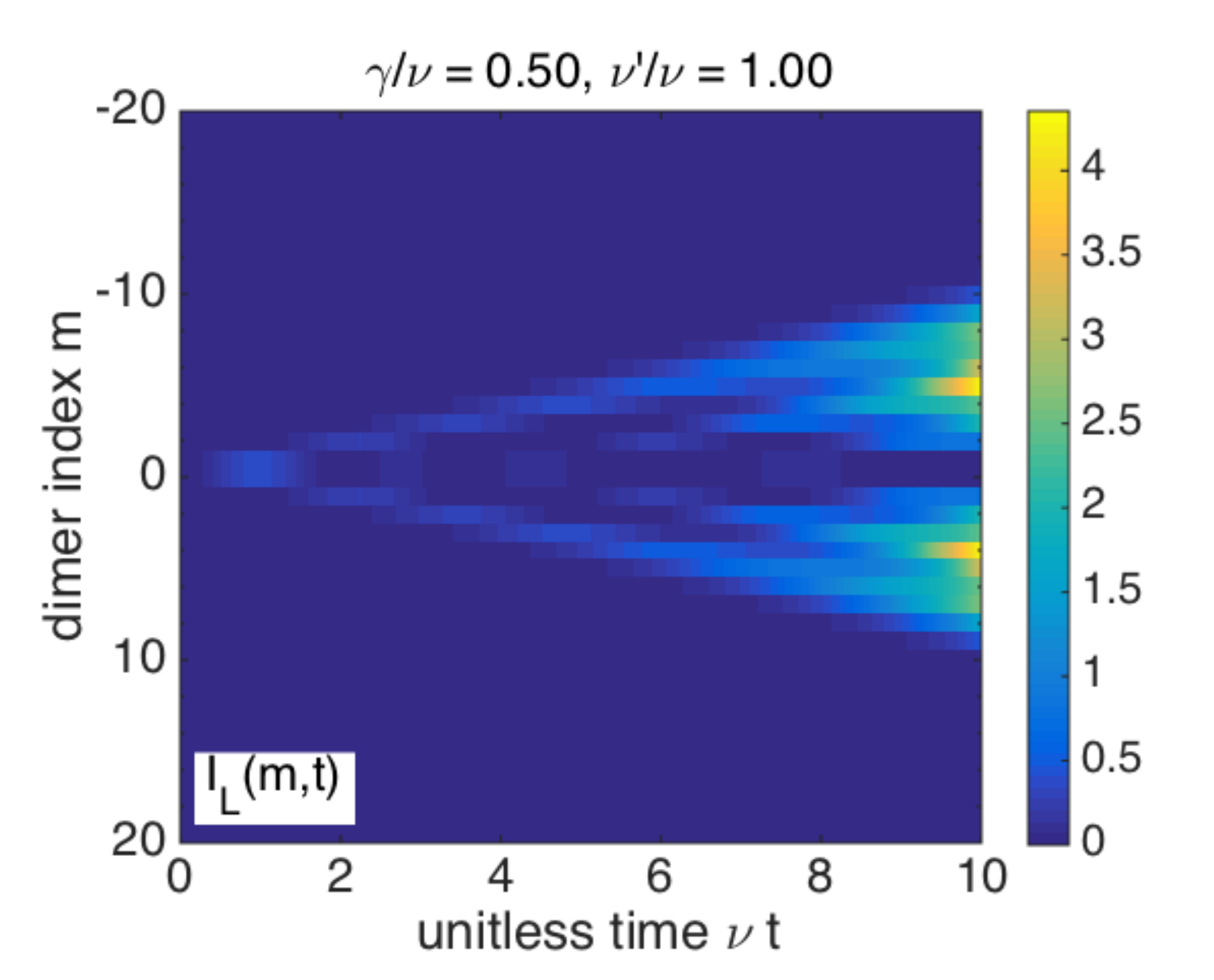}
\label{fig:L-51}}
\caption[]{The gain~\subref{fig:G-51} and loss~\subref{fig:L-51} sublattice intensities for an $N=41$ 
dimer lattice with $\nu'/\nu = 1$ and gain-loss strength $\gamma/\nu=0.5$. The vertical axis shows the dimer index $m$ with $-20\leq m\leq 20$, and the horizontal axis denotes normalized time $\nu t$. Note the order of magnitude difference between intensities on the gain sublattice and the loss sublattice. }
\label{fig:ab}
\end{figure}
In the \PT-symmetric phase, the eigenvalues $\epsilon_k$ are real for all $k$, the non-unitary time evolution generated by the Hamiltonian is periodic, and the total intensity $I(t)=\langle\psi(t)|\psi(t)\rangle$ of an initially normalized wave packet $|\psi(t)\rangle$ remains bounded as a function of time. When the gain-loss strength exceeds $\gamma_{PT}$ but is smaller than $\max_{k}|\nu_k|=\nu+\nu'$, some Fourier components of the initial state grow exponentially while others remain bounded, leading to a total intensity $I(t)$ that oscillates with an amplitude that increases exponentially with time. For $\gamma>|\nu_k|_{k=0}=\nu+\nu'$,  all Fourier components grow exponentially and so does the net intensity. Figure~\ref{fig:ab} show the  intensities for the gain sublattice $I_G(m,t)=|\langle mG|\psi(t)\rangle|^2$ and the lossy sublattice $I_L(m,t)=|\langle mL|\psi(t)\rangle|^2$, for a 41-site dimer lattice with $\nu'/\nu=1$, gain-loss strength $\gamma/\nu=0.5$, and an initial state localized on the gain site of the central dimer, $|\psi(0)\rangle=\delta_{m0}|mG\rangle$. We remind the reader that since the Hamiltonian $H=H_0+V$ is not hermitian, the time-evolved state $|\psi(t)\rangle=\exp(-iHt)|\psi(0)\rangle$  does not have a constant norm. (We use $\hbar=1$.)  We see that the intensities on the two sublattices differ by orders of magnitude; therefore it is useful to consider the two intensity distributions separately.  

Next, we recall the results for the topological transition in a purely lossy dimer lattice~\cite{lr}, and present its generalization to a \PT-symmetric dimer lattice. For a lossy lattice, each dimer has one neutral site ($N$) and one lossy site ($L$). The Hamiltonian for the lossy lattice with open boundary conditions is given by $H^{L}(\nu,\nu',\gamma)=H_0+V^{L}(\gamma)$ where
\begin{equation}
\label{eq:lossy}
V^{L}(\gamma)=-2i\gamma\sum_{m=-M}^{M} |mL\rangle\langle mL|.
\end{equation}
The eigenvalues of the non-hermitian, non-\PT-symmetric Hamiltonian $H^{L}_k$ have a purely decaying part for all eigenmomenta $k$, and therefore any typical initial state is eventually completely absorbed. The mean displacement of the wave packet before it is absorbed is determined solely by 
the intensities on the loss sublattice,
\begin{equation}
\label{eq:deltam}
\Delta m(\nu,\nu',\gamma)=\sum_{m}m \int_{0}^\infty dt\,4\gamma I_L(m,t).
\end{equation} 
Prima facie, Eq.(\ref{eq:deltam}) represented a complicated global measure of the intensity distribution on the loss sublattice; it depends on the initial state, the decay rate $\gamma$, and the two tunneling amplitudes $\nu,\nu'$ that characterize the dimer lattice. For an initial state localized on the neutral site in the central dimer, $m=0$, however, it can be shown - through some non-trivial algebra~\cite{lr} - that the mean displacement $\Delta m$ is equal to the winding number of the $k$-space tunneling amplitude $\nu_k^*=\nu+\nu'\exp(-ik)$~\cite{lr}. Since the winding number is a topological quantity that changes discontinuously and is robust against small disorder perturbations, it follows that the mean displacement, defined by Eq.(\ref{eq:deltam}), is quantized and robust. It changes sharply from 0 to -1 as the inter-dimer tunneling strength $\nu'$ exceeds the intra-dimer tunneling strength $\nu$, and is independent of the decay rate $\gamma>0$.  

Physically, this result can be understood as follows: when the inter-dimer coupling $\nu'$ is small, the wave packet is primarily absorbed on the loss site within the initial dimer; on the other hand, when the inter-dimer coupling $\nu'$ becomes large, the loss-site corresponding to absorption is in the dimer to the left, with index $m=-1$. Although the mean-time to absorption depends on the decay rate $\gamma$, since Eq.(\ref{eq:deltam}) integrates over all possible times, the final result is independent of the decay rate. Being topological in its origin, the analytical result for $\Delta m$ is independent of the loss strength and small disorder, but is valid only for the specific initial state in an infinite lattice~\cite{lr}. Experimentally, the transition is substantially softened and broadened due to the finite size and disorder effects~\cite{julia}. It follows from Fig.~\ref{fig:ptdimerlattice} that a dimer lattice with $\nu'>\nu$ after time reversal and shift by half-a-cell is equivalent to a dimer lattice with $\nu<\nu'$. 

These two lattices - a \PT-symmetric dimer lattice~\cite{kottos} and the purely lossy dimer lattice \'{a} la Rudner and Levitov~\cite{lr} - are equivalent to each other because their respective Hamiltonians differ only by a non-hermitian shift proportional to the identity, $H(\nu,\nu', \gamma)=i\gamma\cdot\mathbf{1}+H^{L}(\nu,\nu',\gamma)$. Therefore, we can define a scaled mean-displacement by considering the intensities on the lossy sublattice~\cite{aay},
\begin{equation}
\label{eq:deltampt}
\Delta m_{PT}(\nu,\nu',\gamma)=\sum_m m\int_{0}^\infty dt\, 4\gamma e^{-2\gamma\,t}I_L(m,t).
\end{equation}

It follows that the scaled mean displacement $\Delta m_{PT}$ undergoes a topological transition at $\nu'=\nu$ which corresponds to a vanishing \PT-breaking threshold, $\gamma_{PT}=|\nu-\nu'|=0$. Therefore, in a \PT-symmetric dimer, the topological transition in $\Delta m_{PT}$ always occurs in the \PT-broken phase. Note that for a general initial state, the intensities $I_G(m,t)$ and $I_L(m,t)$ on both sub-lattices increase exponentially with time in the \PT-symmetry broken phase. However, the integral in Eq.(\ref{eq:deltampt}) converges. In the following section, we numerically investigate the signatures of this transition in the site- and time-dependent intensities $I_G(m,t)$ and $I_L(m,t)$ on the gain and loss sublattices respectively. 

%--------------------------------------------------------------------------------------%

\section{Numerical results}
\label{sec:3}
\begin{figure}
\centering
\subfloat[][Gain sublattice intensity $I_G(m,t)$]{\includegraphics[width=\textwidth]{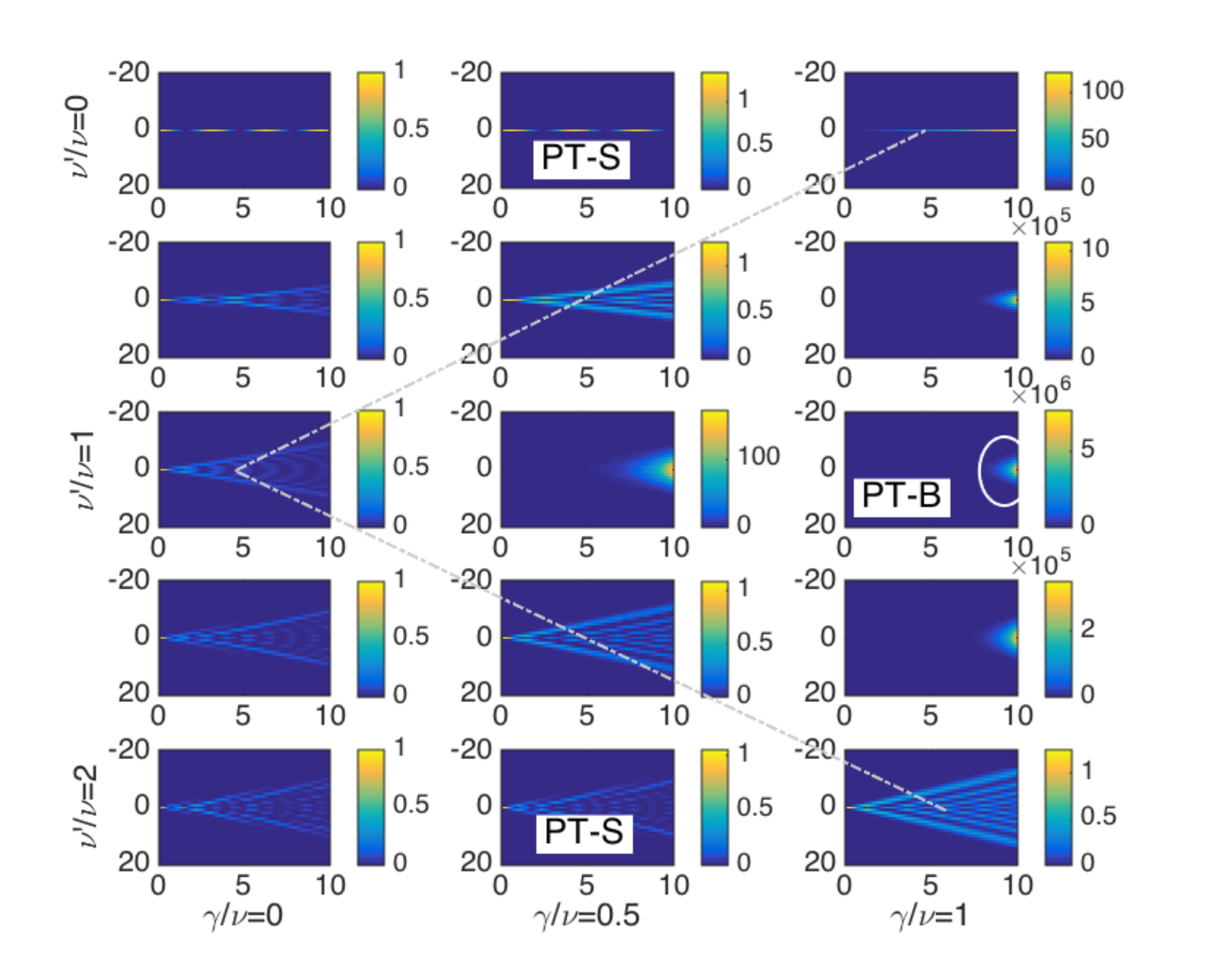}
\label{fig:sweepG1}}\\
\subfloat[][Loss sublattice intensity $I_L(m,t)$]{\includegraphics[width=\textwidth]{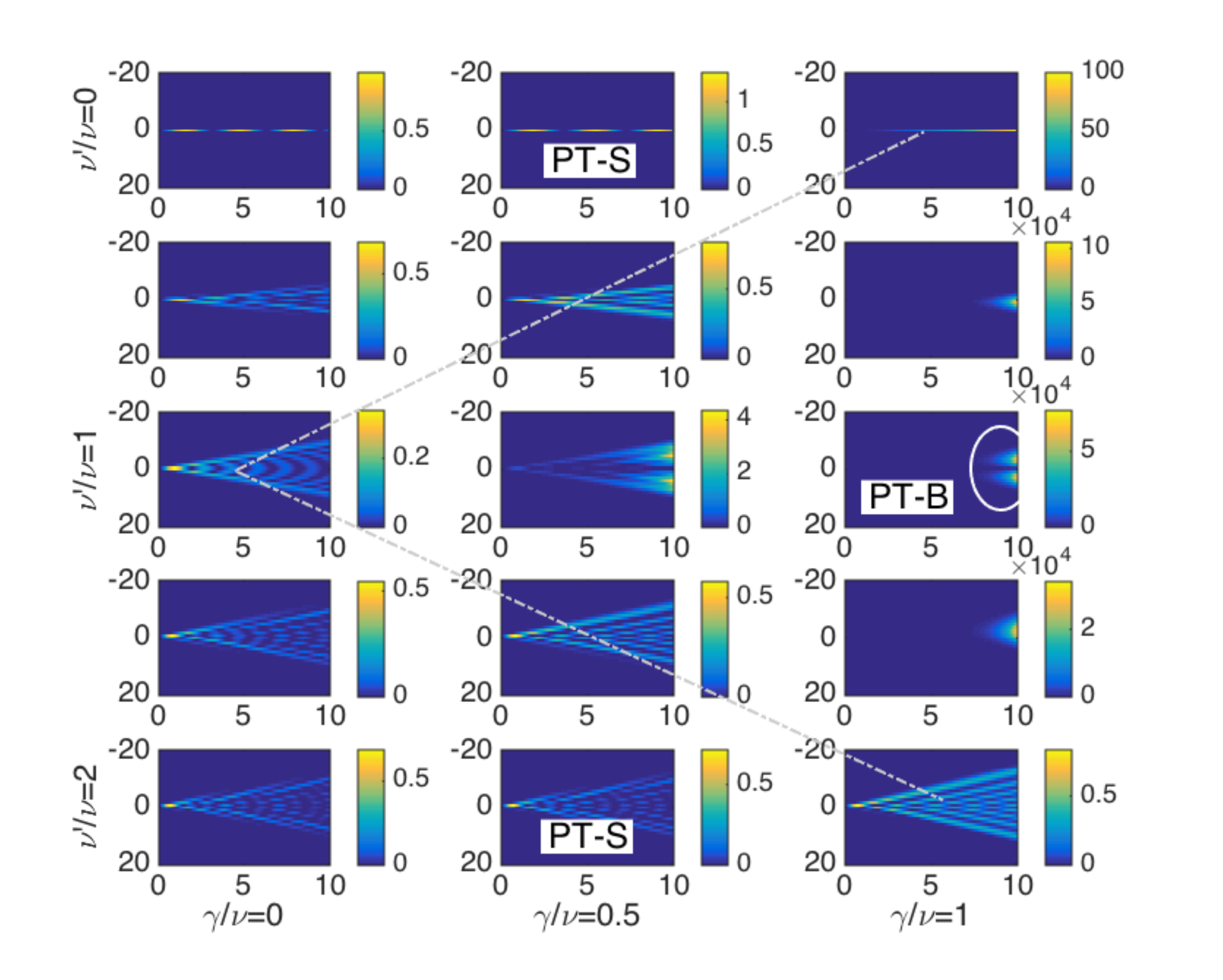}
\label{fig:sweepL1}}
\caption[]{Evolution of gain-sublattice~\subref{fig:sweepG1} and loss-sublattice~\subref{fig:sweepL1} 
intensities for $0\leq\nu'/\nu\leq 2$ and $0\leq\gamma/\nu\leq 1$, and an initial state on the gain-site of the central dimer, $|\psi(0)\rangle=\delta_{m0}|mG\rangle$. Vertical axis in each frame denotes the dimer index $m$ and the horizontal axis denotes normalized time $\nu t$. The dot-dashed gray lines denote the \PT-symmetric phase boundary $\gamma_{PT}/\nu=|1-\nu'/\nu|$. In the \PT-symmetric phase (PT-S), the intensities are bounded and oscillatory. In the \PT-broken phase (PT-B), they are Gaussian except for the loss-sublattice distribution $I_L(m,t)$ at the topological transition $\nu=\nu'$, denoted by a white oval in~\subref{fig:sweepL1}.}
\label{fig:sweepGL1}
\end{figure}

\begin{figure}
\centering
\subfloat[][Loss sublattice intensity $I_L(m,t)$]{\includegraphics[width=\textwidth]{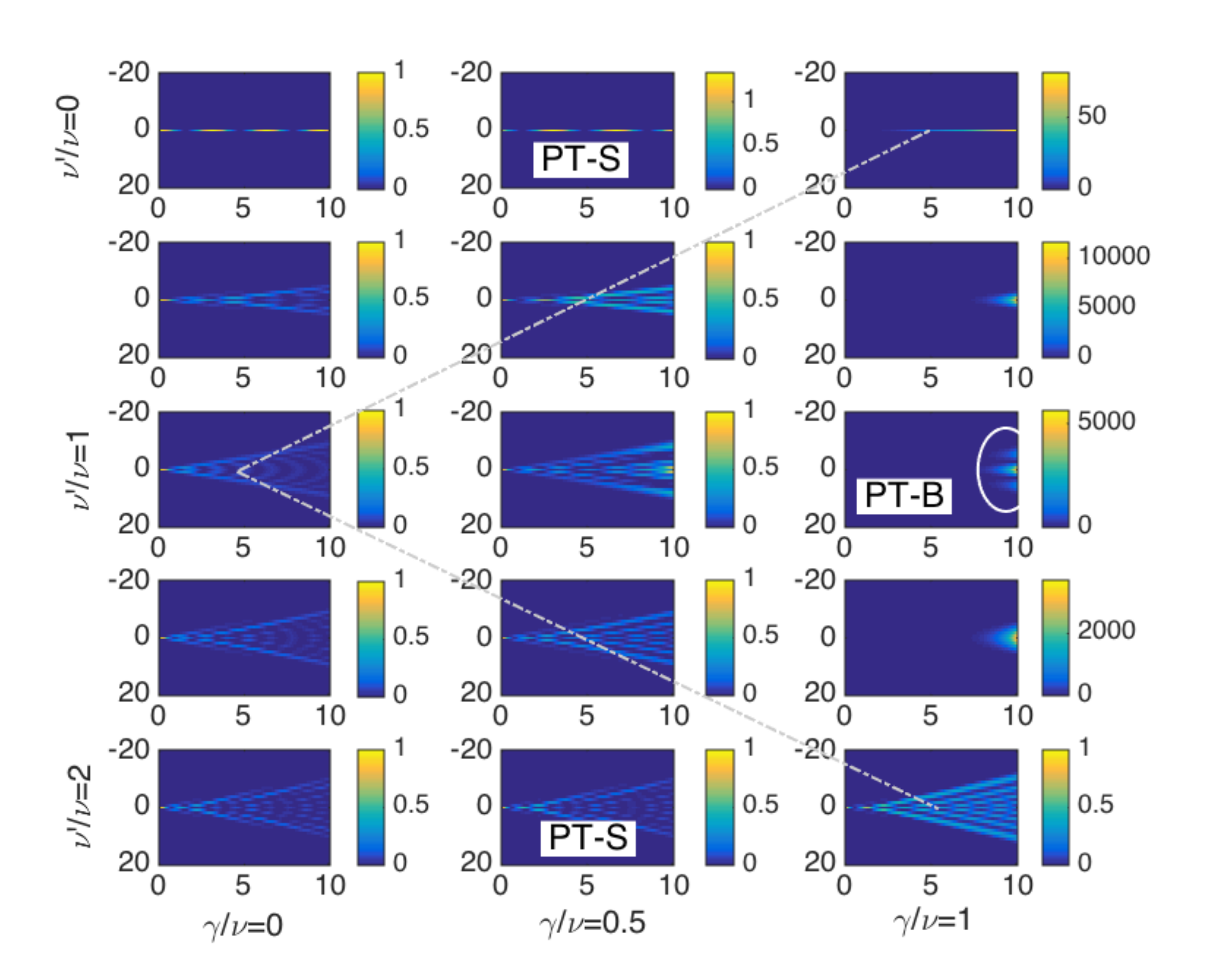}
\label{fig:sweepL0}}\\
\subfloat[][Gain sublattice intensity $I_G(m,t)$]{\includegraphics[width=\textwidth]{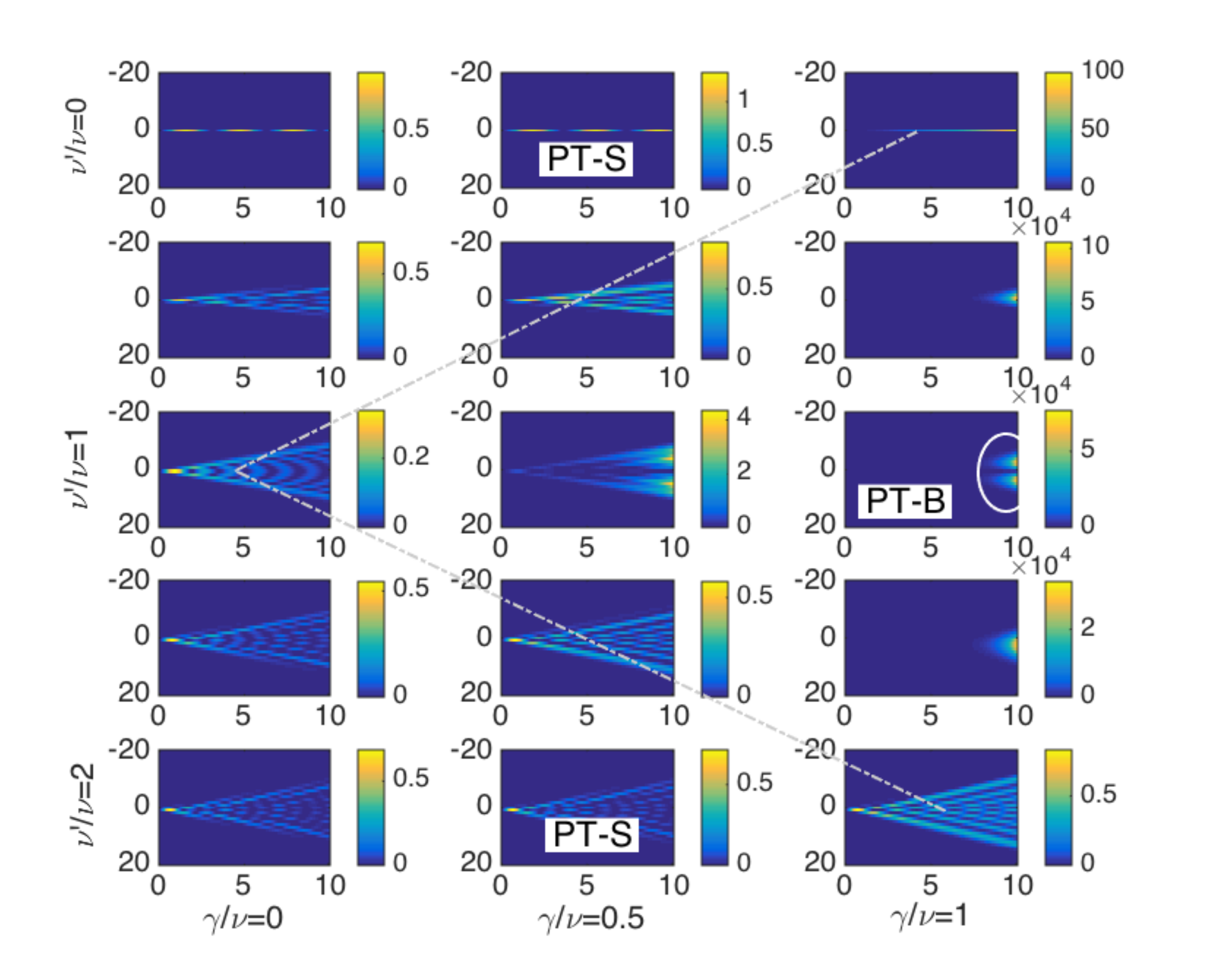}
\label{fig:sweepG0}}
\caption[]{Loss-sublattice~\subref{fig:sweepL0} and gain-sublattice~\subref{fig:sweepG0} intensities for an initial state localized on the loss site, $|\psi(0)\rangle=\delta_{m0}|mL\rangle$. Vertical axis in each frame denotes the dimer index $m$ and the horizontal axis denotes normalized time $\nu t$. The dot-dashed gray lines denote the \PT-symmetric phase boundary $\gamma_{PT}/\nu=|1-\nu'/\nu|$. Both intensities $I_L(m,t)$ and $I_G(m,t)$ show Gaussian behavior except at $\nu'=/\nu$, marked by white ovals in both panels.}
\label{fig:sweepGL0}
\end{figure}

The two independent transitions in the \PT-symmetric dimer lattice are driven by two dimensionless parameters, namely the tunneling ratio $\nu'/\nu$ which governs the topological transition, and the gain-loss strength $\gamma/\nu$ which determines the \PT-breaking phase boundary. Fig.~\ref{fig:sweepGL1} and Fig.~\ref{fig:sweepGL0} show the gain and loss sublattice intensities for $\nu'/\nu=\{0,0.5,1,1.5,2\}$ and $\gamma/\nu=\{0,0.5,1\}$. In each frame, the vertical axis denotes the dimer index $m$ ranging from -20 to 20, and the horizontal axis denotes normalized time $\nu t$ ranging from 0 to 10. Note that when $\nu'/\nu=0$ (top row in all panels), the system consists of uncoupled, \PT-symmetric dimers, and therefore the wave packet remains confined to the central dimer alone; as $\nu'/\nu$ increases, the lateral spread of the wave packet across  the lattice also increases. 

% G-L sublattice intensities for initial state on G-site.
First, let us consider the time evolution of a wave packet initially localized on the gain site of the central dimer, $|\psi(0)\rangle=\delta_{m0}|mG\rangle$. Panel (a) in Fig.~\ref{fig:sweepGL1} shows the gain-sublattice intensity $I_G(m,t)$ and panel (b) shows the corresponding loss-sublattice intensity $I_L(m,t)$. Note that the topological transition occurs across the central row, $\nu'=\nu$, whereas the \PT-breaking transition occurs across the two dot-dashed grey lines, given by $\gamma/\nu=|1-\nu'/\nu|$. Therefore, in both panels, we see that the sublattice intensities are bounded and oscillatory in the \PT-symmetric phase (PT-S). In the \PT-broken phase (PT-B), the gain-sublattice distribution $I_G(m,t)$ shows a single Gaussian whose intensity is maximum at $\nu'=\nu$ because it corresponds to a vanishing \PT-breaking threshold. The loss-sublattice distribution $I_L(m,t)$ also shows a single Gaussian, except at $\nu'=\nu$, when the intensity shows a symmetric, bimodal distribution, marked by the white oval in panel (b).  

% G-L sublattice intensities for initial state on L-site. 
The time-evolution of a state initially localized on the loss-site of the central dimer, $|\psi(0)\rangle=\delta_{m0}|mL\rangle$ is shown in Fig.~\ref{fig:sweepGL0}. Note that in this case, the mean displacement $\Delta m_{PT}$ does not undergo any change as $\nu'/\nu$ is varied; it remains zero, meaning the particle is primarily absorbed on the loss-site it is initially located on~\cite{julia}. The dash-dotted gray lines in both panels denote the \PT-symmetry breaking threshold $\gamma/\nu=|1-\nu'/\nu|$. Both panels show that in the \PT-symmetric phase (PT-S), the time-evolution is oscillatory and the net intensities on the gain and the loss sublattices are comparable to each other. 

Panel (a) shows that in the \PT-broken phase (PT-B), the loss-sublattice intensity profile $I_L(m,t)$ has a {\it symmetric, trimodal distribution}, marked by a white oval in~\subref{fig:sweepL0} when $\nu'/\nu=1$. This is in sharp contrast to the results for $\nu'/\nu\neq 1$, when the distribution consists of a single Gaussian. Panel (b) shows that in the \PT-broken phase, the average gain-site intensity is orders of magnitude higher than the average loss-site intensity. The gain intensity $I_G(m,t)$ shows a symmetric, bimodal distribution exactly at $\nu'/\nu=1$ whereas for $\nu'/\nu\neq 1$, the intensity distribution has a single Gaussian peak. 

Thus, the key numerical observations can be summarized as follows. At $\nu'/\nu=1$, when the winding number of $\nu_k=\nu+\nu'\exp(ik)$ changes from 0 to 1, for an initial state on the gain sublattice, the gain intensity $I_G(m,t)$ shows a single Gaussian peak, whereas the loss intensity $I_L(m,t)$ shows a two-peak structure. When the initial state is localized on the loss sublattice, the gain intensity $I_G(m,t)$ shows a two-peak structure whereas the loss intensity $I_L(m,t)$ shows a structure with three peaks. When $\nu'/\nu\neq 1$, both gain and loss intensities show a single Gaussian peak at long times in the \PT-broken region. In the next section, we will analytically investigate this behavior. 

%--------------------------------------------------------------------------------------%

\section{Analytical approximations in the \PT-broken region}
\label{sec:4}
In this section, we will develop approximate analytical expressions for the real-space, time-dependent wave functions for the two sublattices in the \PT-broken region. As discussed in Sect.~\ref{sec:2}, the \PT-symmetric dimer Hamiltonian is most easily diagonalized in the Fourier space, and the first emergence of complex-conjugate eigenvalues occurs at $k=\pi$. In the \PT-broken phase, the $2\times 2$ time evolution operator is given by~\cite{review}
\begin{equation}
\label{eq:g}
G_k(t)=\exp(-iH_kt)=\cosh(\Gamma_k t)\mathbf{1}-i \frac{H_k}{\Gamma_k} \sinh(\Gamma_k t),
\end{equation}
where $\Gamma_k=\sqrt{\gamma^2-|\nu_k|^2}>0$ is the effective amplification rate. At long times $\Gamma_kt\gg 1$, the Fourier-space time-evolution operator becomes $G_k(t)=\exp(\Gamma_k t)(\mathbf{1}-iH_k/\Gamma_k)/2$. Therefore, equivalently, the real space propagator is given by 
\begin{equation}
\label{eq:psireal}
G_{mn}(t)=\frac{1}{4\pi}\int_{0}^{2\pi} dk\, e^{i(m-n)k+\Gamma_k t}\left(\mathbf{1}-i\frac{H_k}{\Gamma_k}\right).
\end{equation}
Note that Eq.(\ref{eq:psireal}) is valid in the \PT-broken phase even if the eigenvalues of $H_k$ are real for momenta away from $k=\pi$. These momenta, with real eigenvalues $\epsilon_k$, lead to a time evolution operator $G_k(t)$ with bounded norm, and therefore their contribution to Eq.(\ref{eq:psireal}) is vanishingly small at long times $\Gamma_k t\gg 1$. Since the largest contribution to the integral arises from a vanishingly small neighborhood of $k=\pi+p$, the integrand in Eq.(\ref{eq:psireal}) is estimated by approximating $\Gamma_{\pi+p}\approx\Gamma-D p^2/2$ with $\Gamma^2=\gamma^2-\gamma_{PT}^2$ and $D=\nu\nu'/\Gamma$, leading to 
\begin{equation}
\label{eq:psireal2}
G_{m0}(t)=\frac{(-1)^{m} e^{\Gamma t}}{4\pi}\int_{-\infty}^\infty dp\, e^{ipm -D t p^2/2}
\left[\begin{array}{cc} 1+\frac{\gamma}{\Gamma} & -\frac{i}{\Gamma}(\nu-\nu' e^{+ip}) \\
-\frac{i}{\Gamma}(\nu-\nu'e^{-ip}) & 1-\frac{\gamma}{\Gamma- D p^2/2}\end{array}\right].
\end{equation}
Here, without loss of generality, we have chosen $n=0$ as the location of the initial wave packet, and extended the integration range for $p$ to the entire real line because in the long-time limit, $Dt\gg 1$, the integrand contains a Gaussian sharply peaked at $p=0$. We have retained the $p^2$ dependence in the denominator of one of the matrix elements because the matrix element otherwise vanishes at the topological transition boundary $\nu=\nu'$. It is now straightforward to carry out the Gaussian integrals and obtain explicit expressions for the time-dependent wave function at long times in the \PT-broken phase. 

For an initial state localized on the gain-sublattice, $|\psi(0)\rangle=\delta_{n0}|nG\rangle$ (Fig.~\ref{fig:sweepGL1}), we obtain the following expressions for the gain and loss sublattice wave functions,
\begin{eqnarray}
\label{eq:psigg}
\psi_G(m,t)&\sim & \frac{(-1)^m e^{\Gamma t}}{\sqrt{8\pi Dt}}\left(1+\frac{\gamma}{\Gamma}\right)\exp\left[-\frac{m^2}{2D t}\right],\\
\label{eq:psilg}
\psi_L(m,t) &\sim & \frac{i(-1)^{m} e^{\Gamma t}}{\Gamma\sqrt{8\pi Dt}}\left\{\nu\exp\left[-\frac{m^2}{2D t}\right]-\nu'\exp\left[-\frac{(m+1)^2}{2D t}\right]\right\}.
\end{eqnarray}
Note that both wave functions grow exponentially with the amplification rate $\Gamma\leq \gamma$. It follows from Eq.(\ref{eq:psigg}) that the wave function $\psi_G(m,t)$ describes a classical, diffusing particle with diffusion constant $D=\nu\nu'/\Gamma$. This result is expected because, in the \PT-broken phase, where the wave packet intensity increases exponentially with time, we should recover the classical behavior~\cite{diffusive}. For the loss sublattice, we find that $\psi_L(m,t)$ is the difference of two diffusing Gaussians with centers at $m=0$ and $m=-1$ respectively, weighted by the intra-dimer and inter-dimer tunneling strengths. In particular when the topological transition takes place, $\nu=\nu'$, the loss sublattice wave function $\psi_L(m,t)$ shows a symmetric, two-peak structure. 

For an initial state localized on the loss-sublattice, $|\psi(0)\rangle=\delta_{n0}|nL\rangle$ (Fig.~\ref{fig:sweepGL0}), the wave functions are given by 
\begin{eqnarray}
\label{eq:psigl}
\psi_G(m,t) &\sim & \frac{i(-1)^m e^{\Gamma t}}{\Gamma\sqrt{8\pi Dt}}\left\{\nu\exp\left[-\frac{m^2}{2D t}\right]-\nu'\exp\left[-\frac{(m-1)^2}{2D t}\right]\right\},\\
\label{eq:psill}
\psi_L(m,t) &\sim & \frac{(-1)^m e^{\Gamma t}}{\sqrt{8\pi Dt}}e^{-m^2/2D t}\left[1-\frac{\gamma}{\Gamma}\left( 1+\frac{1}{2\Gamma t}-\frac{m^2}{2\nu\nu' t^2}\right)\right].
\end{eqnarray}
It follows from Eq.(\ref{eq:psigl}) that the gain-sublattice intensity distribution is the difference of two diffusing Gaussians centered at $m=0$ and $m=+1$, weighted by the tunneling strengths. In particular, when $\nu=\nu'$, we obtain the symmetric, bimodal distribution seen in panel (b) of Fig.~\ref{fig:sweepGL0}. Eq.(\ref{eq:psill}) implies that the loss-sites wave function $\psi_L(m,t)$ is a diffusing Gaussian centered at $m=0$. However, {\it only at $\nu=\nu'$, the leading order term in the square bracket vanishes}. It generates a multiplicative factor $(1-m^2/D t)$ that accompanies the diffusive Gaussian. This implies that the loss-sublattice intensity vanishes at $m^*=\pm\sqrt(Dt)=\pm\sqrt(\nu^2t/\gamma)$, and gives rise to the three-peak structure seen in panel (a) of Fig.~\ref{fig:sweepGL0}. 

These results can be easily generalized to an arbitrary state on the central dimer, $|\psi(0)\rangle=\delta_{n0}(\cos\theta |nG\rangle + \sin\theta e^{i\phi}|nL\rangle)$. When the initial state has no transverse momentum, $\phi\neq0\mod\pi$, the system does not undergo a topological transition~\cite{aay}, whereas when $\phi=0\mod\pi$, it has a quantized scaled mean displacement. 

%--------------------------------------------------------------------------------------%

\section{Conclusion}
\label{sec:5}
In this paper, we have investigated the interplay between two transitions that are predicted to take place in a \PT-symmetric dimer (or SSH) model.The \PT-symmetry breaking transition is governed by the gain-loss strength $\gamma$ relative to the the tunneling modulation strength $|\nu-\nu'|$, whereas the topological transition in the scaled mean displacement $\Delta m_{PT}$ is governed by the ratio of the inter-dimer to intra-dimer tunneling $\nu'/\nu$. 

We have shown that the gain and loss sublattice intensity profiles, $I_G(m,t)$ and $I_L(m,t)$ respectively, show distinct features at the intersection of the topological transition $\nu=\nu'$ and the \PT-symmetry breaking transition $\gamma_{PT}=|\nu-\nu'|$. These features can be understood through long-time behavior of gain-site and loss-site wave functions, which also capture the classical, diffusive behavior that is expected in the \PT-broken phase. Although experimental realization of an active SSH model, where half the waveguides have a constant amplification, is challenging, it is feasible with the current sample fabrication technology; therefore, we expect that all of its attendant properties, including symmetric, edge-localized states will be observable in it. 

In this work, we have not considered the effects of nonlinearity~\cite{igor}. In the \PT-broken phase, the nonlinearity manifests itself in two ways. First, it introduces a state-dependent potential $V_G(m)\propto |\psi_G(m,t)|^2$ on each gain site and a corresponding potential $V_L(m)$ on each loss site; physically, this potential represents the intensity-dependent change in the local index of refraction~\cite{boris1,boris2,dnc}. Second, as the site-dependent intensity increases, the model with {\it constant, local-intensity independent} gain and loss coefficients becomes less reliable~\cite{lasers}. Thus, our findings are valid in a range of parameters where the effects of nonlinearity are mitigated. They suggest that the interplay between \PT-symmetry breaking transition, and topological transitions in one or two dimensional \PT-symmetric models leads to interesting results. 

%--------------------------------------------------------------------------------------%

\begin{acknowledgement}
The authors thank Avadh Saxena for useful discussions. This work was supported by NSF DMR-1054020. 
\end{acknowledgement}

%--------------------------------------------------------------------------------------%

\bibliographystyle{spphys}

%--------------------------------------------------------------------------------------%
	
\end{document}